\documentclass{aastex}
\usepackage{emulateapj5}

\def\ts     {\thinspace}
\def\kms    {\ifmmode{{\rm \ts km\ts s}^{-1}}\else{\ts km\ts s$^{-1}$}\fi}
\def\msol     {\ifmmode{{\rm M}_{\odot}}\else{M$_{\odot}$}\fi} 
\def\etal   {{\rm et\ts al.}}
\def\aco  {\ifmmode{^{12}{\rm CO}(J\!=\!1\! \to \!0)}\else{$^{12}{\rm CO}(J\!=\!1\! \to \!0)$}\fi}
\def\bco  {\ifmmode{^{12}{\rm CO}(J\!=\!2\! \to \!1)}\else{$^{12}{\rm CO}(J\!=\!2\! \to \!1)$}\fi}
\def\m    {\ifmmode{\mu {\rm m}}\else{$\mu$m}\fi}
\def\cco  {\ifmmode{^{13}{\rm CO}(J\!=\!1\! \to \!0)}\else{$^{13}{\rm CO}(J\!=\!1\! \to \!0)$}\fi}
\def\dco  {\ifmmode{^{13}{\rm CO}(J\!=\!2\! \to \!1)}\else{$^{13}{\rm CO}(J\!=\!2\! \to \!1)$}\fi}
\def\eco  {\ifmmode{{\rm C}^{18}{\rm O}(J\!=\!1\! \to \!0)}\else{${\rm C}^{18}{\rm O}(J\!=\!1\! \to \!0)$}\fi}
\def\nh   {\ifmmode{N(\hi)}\else{$N$(\hi)}\fi}
\def\hun    {\ifmmode{I_{100}}\else{$I_{100}$}\fi}
\def\sex    {\ifmmode{I_{60}}\else{$I_{60}$}\fi}
\def\hh     {\ifmmode{{\rm H}_2}\else{H$_2$}\fi}
\def\nhh     {\ifmmode{N({\rm H}_2)}\else{$N$(H$_2$)}\fi}
\def\mhh     {\ifmmode{M({\rm H}_2)}\else{$M$(H$_2$)}\fi}
\def\zwco   {\ifmmode{^{12}{\rm CO}}\else{$^{12}{\rm CO}$}\fi}
\def\nzwco   {\ifmmode{N(^{12}{\rm CO})}\else{$N(^{12}{\rm CO})$}\fi}
\def\wzwco   {\ifmmode{W(^{12}{\rm CO})}\else{$W(^{12}{\rm CO})$}\fi}
\def\drco   {\ifmmode{^{13}{\rm CO}}\else{$^{13}{\rm CO}$}\fi}
\def\ndrco   {\ifmmode{N(^{13}{\rm CO})}\else{$N(^{13}{\rm CO})$}\fi}
\def\wdrco   {\ifmmode{W(^{13}{\rm CO})}\else{$W(^{13}{\rm CO})$}\fi}
\def\tex    {\ifmmode{T_{ex}({\rm CO})}\else{$T_{ex}({\rm CO})$}\fi}
\def\ha     {\ifmmode{{\rm H}\alpha}\else{${\rm H}\alpha$}\fi}
\def\amm     {\ifmmode{{\rm NH}_{3}}\else{${\rm NH}_{3}$}\fi}
\def\xco     {\ifmmode{X_{\rm CO}}\else{$X_{\rm CO}$}\fi}
\def\jyk     {\ifmmode{{\rm Jy\,km\ts s}^{-1}}\else{Jy\,km\ts s$^{-1}$}\fi}

\newcommand{\hi}{H\,{\small{\sc I}}}

\slugcomment{}

\shorttitle{Molecular Outflow and Streamers in M\,82}
\shortauthors{Walter, Wei\ss\ \& Scoville}

\begin{document}

\title{Molecular Gas in M\,82: Resolving the Outflow and Streamers}

\author{Walter F.}
\affil{Owens Valley Radio Observatory, California Institute of Technology, 
Pasadena, CA, 91125}
\email{fw@astro.caltech.edu}

\author{Wei\ss\ A.}
\affil{Radioastronomisches Institut der Universit\"at Bonn, 
Auf dem H\"ugel 71, 53121 Bonn, Germany\\
IRAM, Avenida Divina Pastora 7, NC, 18012 Granada, Spain}
\email{aweiss@astro.uni-bonn.de}

\and

\author{Scoville N.}
\affil{Astronomy Department, California Institute of Technology, 
Pasadena, CA, 91125}
\email{nzs@astro.caltech.edu}

\begin{abstract}

We present a high--resolution ($3.6''$, 70\,pc) \aco\ mosaic of the
molecular gas in M\,82 covering an area of $2.5'\times3.5'$
(2.8\,kpc$\times$3.9\,kpc) obtained with the OVRO millimeter
interferometer. The observations reveal the presence of huge amounts
of molecular gas ($>$ 70\% of the total molecular mass, M$_{\rm
tot}\approx1.3\times10^9$\msol) outside the central 1\,kpc disk.
Molecular streamers are detected in and below M\,82's disk out to
distances from the center of $\sim$1.7\,kpc. Some of these streamers
are well correlated with optical absorption features; they form the
basis of some of the prominent tidal HI features around M\,82. This
provides evidence that the molecular gas within M\,82's optical disk
is disrupted by the interaction with M\,81.  Molecular gas is found in
M\,82's outflow/halo, reaching distances up to 1.2\,kpc below the
plane; CO line--splitting has been detected for the first time in the
outflow. The maximum outflow velocity is $\sim 230$\,km\,s$^{-1}$; we
derive an opening angle of $\sim55^{\circ}$ for the molecular outflow
cone.  The total amount of gas in the outflow is
$>3\times10^8$\,\msol\ and its kinetic energy is of order
$10^{55}$\,erg, about one percent of the estimated total
mechanical energy input of M\,82's starburst.  Our study implies that
extreme starburst environments can move significant amounts of
molecular gas in to a galaxy's halo (and even to the intergalactic
medium).

\end{abstract}
\keywords{galaxies:individual (M\,82)--galaxies: interactions--galaxies: 
kinematics and dynamics--galaxies:ISM--galaxies: starburst--ISM: jets
and outflows}

\section{Introduction}

M\,82 is the prototypical nearby starburst galaxy; innumerous studies
of M\,82 have been performed in the last decades. Most of these
studies focused on the extreme starburst properties and the related
prominent outflow emerging from the inner 1\,kpc of M\,82 (at a
distance of 3.9\,Mpc, 1$'\sim1.1$\,kpc; Sakai \& Madore 2001).  The
tremendous star forming activity in M\,82's central region can be
traced by the high FIR luminosity (e.g., Telesco \& Harper, 1980), the
filamentary H$\alpha$ outflow (e.g., Bland \& Tully, 1988; Devine \&
Bally, 1999), the extended X--ray outflow (e.g., Strickland, Ponman \&
Stevens, 1997) and the existence of $\sim$\,20 compact radio sources
identified as supernova (SN) remnants and compact HII regions (e.g.,
Kronberg et al., 1981, Pedlar \etal\ 1999).  The central starburst is
believed to be fueled by the large amount of molecular gas present in
M\,82's center (M$_{\rm tot}\approx2.3\times10^8$\msol, e.g., Wei\ss\
\etal\ 2001).

This central concentration of molecular gas is very strong in CO
emission and has been the subject of many studies. Most of these
studies were done with single dish telescopes (e.g., Stark \& Carlson,
1982; Young \& Scoville, 1984; Wild \etal\ 1992; Mao \etal\ 2000;
Seaquist \& Clark, 2001; Taylor, Walter, \& Yun, 2001) addressing both
the excitation conditions and the large scale distribution of the
molecular gas. Some of these studies already provided evidence for
extended CO emission around M\,82's center. Interferometric CO studies
of M\,82's molecular component have so far been limited to the central
kpc (Shen \& Lo 1993, Neininger \etal\ 1998, Wei\ss\ \etal\ 1999,
Wei\ss\ \etal\ 2001). In this letter, we present the first results of
our extensive mapping campaign of the central
$2.8$\,kpc$\times3.9$\,kpc of M\,82 at a resolution of $3.6''$
($\sim$70\,pc).

\section{Observations}

We observed M\,82 in the CO(1$\to$0) transition using Caltech's Owens
Valley Radio Observatory's millimeter array (OVRO) in C, L and H
configurations (10 pointing mosaic). This mosaic covers a
$2.5'\times3.5'$ field, including a significant fraction of the
optical disk of M\,82 as well as its prominent \ha\ outflow. The
central pointing was centered near the 2.2\m\ peak at $\alpha =
09^h55^m52^s.2, \delta = 69^{\circ}40'47.1''$ (J2000.0).  The other 9
pointings were shifted by $30''$ (OVRO primary beam: $60''$ at
115\,GHz), the field of view is indicated in Fig.~1 (right). Data were
recorded using a correlator setup with 240\,MHz bandwidth and 2\,MHz
resolution, resulting in a velocity resolution of 5.2 \kms\ with a
total bandwidth of 600 \kms\ at 115\,GHz.

For image generation and processing we used the {\sc miriad} software
package.  Because extended structure was missed in our interferometer
data, the data cube was zero--spacing corrected using the IRAM 30\,m
data described in Wei\ss\ \etal\ (2001). The combination was done by
replacing visibilities shorter than the shortest baseline in the final
processed data cube by visibilities computed from the single dish data
(this technique is described in more detail in Wei\ss\ \etal\ 2001).
The final data cube presented here has a spatial resolution of
$3.6''\times 3.6''$ and an rms noise level of 25\,mJy\,beam$^{-1}$
(0.18\,K) in a 5.2 \kms\ channel.

\section{Data Presentation and Analysis \label{results}}

\subsection{Global properties}

The integrated intensity distribution of our combined mosaic is shown
in Fig.~1. The physical properties (CO flux, molecular mass, linear
size) of the newly detected features are summarized in Tab.~1. In
converting the observed CO luminosity to H$_2$ column densities we
applied two conversion (`$X$') factors 1) $X_{\rm M\,82}\,=\nhh/{\rm
I(CO)}=0.5\times10^{20}\,{\rm cm^{-2}\,(K\kms)^{-1}}$, an average
value for M\,82's central disk (Wei\ss\ \etal\ 2001), and 2) $X_{\rm
Gal}\,=1.6\times10^{20}\,{\rm cm^{-2} \,(K
\kms)^{-1}}$, the Galactic conversion factor (Hunter
\etal\ 1997) which is presumably a more realistic value for a
non--starburst environment (such as the molecular streamers S1-S4, see
Fig.~1, right). The total flux detected in the field of view is
17680\,\jyk\ (estimated error: $\sim10\%$). Using $X_{\rm M\,82}$ for
the disk and the outflow O-S\,+\,O-N (which presumably consists of
warm gas emerging from the disk, see Fig.~1, right) and $X_{\rm Gal}$
for the molecular streamers S1--S4, we derive a total molecular mass
of M\,82 of $\sim 1.3\times10^9$\msol\ (total masses using other
conversion factors can be easily derived from Tab.~1).

\subsection{Seperating the CO Components in M\,82}

The different CO emitting components in M\,82 (disk, streamers,
outflow/halo) have clear kinematic signatures which can be seperated
by inspecting consecutive channel maps. We used this velocity
information to subdivide the full OVRO CO data cube into three
sub--cubes out of which individual integrated intensity maps were
created for the disk, streamer and outflow/halo component. These
components are shown in Fig. 3 (overlaid as countours on an H$\alpha$
map by Shopbell et al.\ 1998): The yellow contours represent M\,82's
central disk only (see Sec.~3.3); streamers S1--S4 are shown in orange
(Sec.~3.4), and the red contours represent an intensity map of the
outflow/halo (Sec.~3.5).

\subsection{The Central Molecular Disk}
The central 1\,kpc of M\,82's CO distribution is dominated by the well
known triple--peaked molecular distribution (Fig.~1, see also yellow
contours in Fig.~3) which has been discussed in previous
high--resolution CO studies.  This region coincides with the strong
central star formation and forms the base of the prominent outflow
visible in \ha\ and X--rays. It is striking that only about a third of
M\,82's total CO {\em flux} is associated with the center (see
Tab.~1).  Depending on the employed X--factor this corresponds to a
molecular {\em mass} fraction of $<30\%$ for the central molecular
disk.

\subsection{Molecular Streamers}
Several molecular streamers are detected west, east and south--east of
the central molecular concentration (labeled S1 to S4 in Fig.~1, see
also orange contours in Fig.~3).  The most prominent streamer is
located west of the molecular disk (S1). Its orientation is shaped in
a similar fashion and has the same velocity as the prominent \hi\
streamer which points towards M81 (the M81--M\,82 tidal tail, streamer
`S' in Yun, Ho \& Lo 1993).  A weaker streamer forms the extension of
the molecular disk towards the north--east (S2), roughly following
another large--scale \hi\ streamer (streamer `N' in Yun, Ho \& Lo
1993). S1 and S2 follow very closely the dust absorption features seen
in optical images (Fig.~2). Two additional molecular streamers are
visible south--east (S3) and east (S4); in contrast to S1 and S2,
these two structures are situated {\em away} from the optical plane of
the galaxy (see also Figs.~1, 3 and~5, bottom). Based on the available
data, we can not rule out a potential interaction between S3/S4 and
the outflow/halo gas.  A summary of the fluxes of S1-S4 is given in
Tab.~1.

\subsection{Molecular Gas in the Outflow/Halo}

{\em Masses:} Molecular gas is clearly associated with the outflow
south (O--S) and north (O--N) of the central molecular disk. The
molecular gas in the outflow/halo is visible up to 0.8\,kpc north and
1.2\,kpc south of the disk.  The total mass of molecular gas in the
outflow/halo accounts to an impressive amount of
$3.3\times10^{8}$\,\msol, even employing the lower conversion factor
$X_{\rm M82}$. 

{\em Outflow morphology:} Interestingly, the CO emission in the
outflow/halo (Fig.~3, red contours) has an almost spherical shape with
a diameter of $\sim1.5\,$kpc (this is similar in size to the dust halo
discussed in Alton et al.\ 1999). In addition, there is a striking
correlation between the H$\alpha$ outflow and the molecular gas in the
south. The correlation is less prominent in the north--east of the
outflow but clearly visible in the north--west. The chimney of shocked
molecular gas detected in SiO by Garc{\' i}a-Burillo et al.\ (2001)
can not be clearly identified in our CO data.

{\em Kinematics:} For the first time CO line splitting is detected in
the outflow of M\,82.  Fig.~4 shows a pV diagram of the CO gas along
the outflow axis of M\,82. Here, a position angle of 150$^{\circ}$
was chosen to allow a comparison with published H$\alpha$ velocities
(see caption). The CO line splitting is clearly visible between
$-20''$ and $-35''$ below the disk; the velocities agree well with the
H$\alpha$ velocities. The blue--shifted CO component is not detected
below $-35''$ but the other component is traced at similar velocities
as the H$\alpha$ up to $-70''$. At $-35''$ (660\,pc below the plane),
the CO is blueshifted by $\sim 130$\kms with respect to the systemic
velocity (v$_{\rm sys}$=220\kms). Correcting for M\,82's inclination
(80$^{\circ}$, de~Vaucouleurs \etal\ 1991) and our derived cone angle
of 50$^{\circ}$ (see below) we derive a deprojected maximum outflow
velocity of $\sim 130$\kms/sin$(35^{\circ})\approx230\kms$. The
situation is more complex in the northern part of the outflow -- here
the molecular gas seems to be entrained in the H$\alpha$ outflow. As
is obvious from Fig.~4, most of the projected velocities are lower
than this value and we estimate an average outflow velocity of the gas
of order $\sim100$\,\kms. Based on the escape velocities presented in
Martin (1998, Figure~9) we estimate that of order 10\% of the outflow
gas ($3.3\times10^{7}$\,\msol, i.e., a few percent of the total gas
mass) may be lost to the intergalactic medium (IGM).

{\em Outflow cone angle:} Fig.~5 presents pV diagrams along and
parallel to the major axis. In the north, the rotation curve is
flattened which can be interpreted as conservation of angular momentum
in the outflow. A similar morphology is seen towards the southern pV
diagram as well; however, this region is clearly disturbed by the
features S3 and S4.

To determine the orientation of the outflow cone, we derived
`flow--lines' of the molecular gas (as first proposed by Seaquist \&
Clark 2001). The basic assumption is that the angular momentum of the
gas in the outflow is conserved. For each distance from the plane $z$
we calculated the angular momentum $L\sim r\cdot v_{\rm t}$ where
$v_{\rm t}$ is the terminal velocity at distance $r$ from the minor
axis. We took $v_{\rm t}$ from our first moment map of our `outflow
cube' (see above); these values give a reasonable approximation for
the true $v_{\rm t}$ (see also the discussion in Seaquist \& Clark
2001).  The result of this analysis is plotted as contours in Fig.~6
-- each contour represents an iso--angular momentum (`$L$') line and
describes the flow--line of an individual particle in the outflow. It
is striking how similar the $L$--contours (derived entirely from {\em
kinematic} information) are compared to the morphology of the ionized
gas in the outflow (e.g., Fig.~3). From the third contour (counting
from the minor axis) in Fig.~6, we derive an average cone angle of
$\sim55^{\circ}$ (north: $\sim60^{\circ}$, south: $\sim50^{\circ}$),
similar to what has been found in previous optical studies (optical:
$\sim30-35^{\circ}$, Bland \& Tully 1988, McKeith \etal\ 1995, to
$\sim60-65^{\circ}$, Heckman, Armus \& Miley 1990). We note that
Seaquist \& Clark (2001) derived an angle of $\sim40^{\circ}$ for the
northern and $\sim90^{\circ}$ for the southern molecular outflow; the
discrepancy in the south can be explained by confusion of their
velocity field by S3 and S4 in their low--resolution data.

{\em Energetics:} To calculate an order of magnitude estimate of the
kinetic energy of the outflow, we assume that the molecular gas in the
outflow ($3.3\times10^8$\msol) has an average outflow velocity of
$\sim 100$\,\kms\ (see discussion above). This gives a kinetic energy
of $\sim 3.3\times10^{55}$\,erg. Extrapolating the current SN rate in
M\,82 of 0.1\,yr$^{-1}$ (Kronberg \etal\ 1981) back to the age of the
current starburst ($5\times10^7$\,yr, O'Connell \etal\ 1995) yields a
total mechanical input by SNe of $\sim 5\times 10^{57}$\,erg. Assuming
an average outflow velocity of 100\,\kms\ and a distance of
$\sim1$\,kpc it takes of order $1\times10^7$\,yr to build up the
molecular outflow, similar to the age of the starburst. This implies
that of order $1\%$ of the released mechanical energy of M\,82's
central starburst has been converted to kinetic energy in the
molecular outflow gas.

\section{Concluding Remarks}

Our high--resolution OVRO observations of the prototypical starburst
galaxy M\,82 provide new exciting insights on the properties of the
molecular gas around the galaxy's center.  The newly detected
molecular streamers (both in and below the galaxy's plane) suggest
that the molecular gas in the inner few kpc is severely affected by
the interaction with M\,81. This redistribution of molecular gas is
likely the trigger for the strong starburst activity in M\,82's
center. An impressive amount of molecular gas is detected in the halo
of M\,82.  It is striking that most of the outflow gas is distributed
in an almost spherical halo; at large distances from the plane its
morphology looks remarkably similar to that of the ionized gas.  The
gas in the halo is rotating slower than in the disk; we find that the
CO kinematics in the halo can be used to constrain the outflow cone
angle of the molecular gas ($\sim55^{\circ}$). Our estimated CO
outflow velocities imply that a significant amount of molecular gas is
not only transported to the halo but may also escape the galactic
potential.  This has important implications for the chemical evolution
of the starburst host galaxy and for the chemical enrichment of the
IGM, especially in the early universe when much more violent
starbursts took place.

\acknowledgments{FW acknowledges NSF grant AST 96-13717. AW acknowledges 
DFG grant SFB~494. We thank P. Shopbell for providing us with his
M\,82 Fabry--Perot H$\alpha$ data.}

\begin{figure}
\begin{center}
\caption{{\em Left:} Logarithmic representation of the 
integrated CO(1$\to$0) map of the zero--spacing corrected OVRO mosaic
(moment 0). Contours are plotted at 1.5, 3, 6, 12, 24, 48, 96, 192
Jy\,beam$^{-1}$\,km\,s$^{-1}$; the rms is
0.3\,Jy\,beam$^{-1}$\,km\,s$^{-1}$.  The OVRO mosaicked field of view
is indicated by the outer envelope (plotted at $1.5\times$ the FWHM of
the primary beam). {\em Right:} Same plot with superimposed labels of
the molecular streamers (S1--S4) and the outflow gas (O--N and O--S);
the lines indicate the orientations of the position velocity diagrams
discussed in Fig.~4 and~6. The dashed lines indicate the derived
opening angle of the molecular outflow (50$^{\circ}$, south;
60$^{\circ}$, north, see Sec.~3.5).}
\end{center}
\end{figure}

\begin{figure}
\begin{center}
\caption{Three color composite of M\,82: {\em Red:} Optical image of 
M\,82; {\em Blue:} H$\alpha$ image of M\,82 (see Fig.~3); {\em
Green:} OVRO CO map (see Fig.~1).}
\end{center}
\end{figure}

\begin{figure}
\begin{center}
\caption{H$\alpha$ image of M\,82 (greyscale, see text). {\em 
White contours:} M\,82 disk, {\em Grey contours:} M\,82 streamer
(S1--S4), {\em Black contours:} molecular gas associated with the
outflow (O--S\,+\,O--N). Contour levels are the same as in Fig.~1, a
contour of 0.4 Jy\,beam$^{-1}$\kms\ has been added to the outflow
component. Note that molecular gas is clearly associated with M\,82's
prominent outflow of ionized gas.}
\end{center}
\end{figure}

\begin{figure}
\begin{center}
\caption{CO pV diagram along the outflow in M\,82 
(PA=150$^{\circ}$, see `pV-O' in Fig.~1 for orientation).  The crosses
(Shopbell \etal\ 1998) and pluses (McKeith \etal\ 1995) represent the
H$\alpha$ velocities. Towards the south (negative offsets), CO line
splitting is detected.}
\end{center}
\end{figure}

\begin{figure}
\begin{center}
\caption{CO pV diagram along and parallel to the major 
axis of M\,82 (PA=75$^{\circ}$, see labels `pV-C', `pV-N' and `pV-S'
in Fig.~1 for the relative orientations, offsets are given with respect
to the minor axis).  Contours in the south are shown at 0.08, 0.15,
0.2, 0.25 Jy\,beam$^{-1}$, along the major axis contours are also
plotted at 0.75, 1.5Jy\,beam$^{-1}$. Note that the rotational amplitude
is decreasing with distance from M\,82's plane. The southern diagram
(bottom) includes part of the streamers S1, S3 and S4.}
\end{center}
\end{figure}

\begin{figure}
\begin{center}
\caption{Iso--angular momentum (`$L$') contours (black) of molecular 
gas in M\,82 (greyscale: moment 0 map, same as in Fig.~1). Under the
assumption that angular momentum in the outflow is conserved, these
contours show the `flow--lines' of the molecular gas (see text).}
\end{center}
\end{figure}

{\begin{table*}[h]
\small{
\begin{center}
\caption{Observed CO fluxes in M\,82}
\begin{tabular}{l r c c c}
\hline
\hline
        & I$_{\rm CO}$ & \multicolumn{2}{c}{\mhh\ [10$^8$\msol]}          & size\\
        & [\jyk]       & X$_{\rm M82}$=0.5$^{\rm a}$ & X$_{\rm Gal}$=1.6$^{\rm a}$ & [kpc]\\
\hline
disk    & 4950 & 2.3  & (7.4) & 1.0\\
 \hline		 
S1      & 2230 & 1.0  & 3.3 & 1.5\\
S2      & 1090 & 0.5  & 1.6 & 1.7\\
S3      & 1510 & 0.7  & 2.2 & 1.2\\
S4      &  670 & 0.3  & 1.0 & 0.9\\
S1--S4  & 5500 & 2.5  & 8.0 & -- \\
\hline		  
O--N    & 3410 & 1.6  & 5.0 & 0.8 \\
O--S    & 3830 & 1.8  & 5.6 & 1.2 \\
O--N\,+\,O--S   & 7240 & 3.3  & 10.6 & -- \\
\hline	
total   & 17680& \multicolumn{2}{c}{13.5$^{\rm b}$} & -- \\
\hline
\hline
\end{tabular}\\
\end{center}
\footnotesize{a: units are: $10^{20}\,{\rm cm^{-2} \,(K \kms)^{-1}}$}\\
\footnotesize{b: disk \& outflow: X$_{\rm M82}$; streamer: X$_{\rm Gal}$ (see text)}\\
}
\end{table*}}

\end{document}